\documentstyle[eqsecnum,aps]{revtex}

\begin{document}
\draft \preprint{HEP/123-qed}
\title{ Path Integral of the Two Dimensional Su-Schrieffer-Heeger Model}
\author{Marco Zoli}
\address{Istituto Nazionale Fisica della Materia - Dipartimento di Fisica,
Universit\'a di Camerino, \\
62032 Camerino, Italy. e-mail: marco.zoli@unicam.it }

\date{\today}
\maketitle
\begin{abstract}
The equilibrium thermodynamics of the two dimensional
Su-Schrieffer-Heeger Model is derived by means of a path integral
method which accounts for the variable range of the electronic
hopping processes. While the lattice degrees of freedom are
classical functions of time and are integrated out exactly, the
electron particle paths are treated quantum mechanically. The free
energy of the system and its temperature derivatives are computed
by summing at any $T$ over the ensemble of relevant particle paths
which mainly contribute to the total partition function. In the
low $T$ regime, the {\it heat capacity over T} ratio shows an
upturn peculiar of a glassy like behavior. This feature is more
sizeable in the square lattice than in the linear chain as the
overall hopping potential contribution to the total action is
larger in higher dimensionality.
\end{abstract}
 \narrowtext
\section*{I. Introduction}

The lattice dimensionality is a key parameter in materials
science. One dimensional (1D) systems with half filled band
undergo a structural distortion \cite{peierls} which increases the
elastic energy and opens a gap at the Fermi surface thus lowering
the electronic energy. The competition between lattice and
electronic subsystems stabilizes the 1D structure which
accordingly acquires semiconducting properties whereas the
behavior of the 3D system would be metallic like. Conjugated
polymers, take polyacetylene as prototype, show anisotropic
electrical and optical properties \cite{lu} due to intrinsic
delocalization of $\pi$ electrons along the chain of CH units. As
the intrachain bonding between adjacent CH monomers is much
stronger than the interchain coupling the lattice is quasi-1D.
Hence, as a result of the Peierls instability, polyacetylene shows
an alternation of short and long neighboring carbon bonds, a
dimerization, accompanied by a two fold degenerate ground state
energy. The Su-Schrieffer-Heeger (SSH) model Hamiltonian
\cite{ssh} has become a successful tool in polymer physics as it
hosts the peculiar ground state excitations of the 1D conjugated
structure and it accounts for a broad range of polymer properties
\cite{ssh1}. As a fundamental feature of the SSH Hamiltonian the
electronic hopping integral linearly depends on the relative
displacement between adjacent atomic sites thus leading to a
nonlocal {\it e-ph} coupling with vertex function depending both
on the electronic and the phononic wave vector. The latter
property induces, in the Matsubara formalism \cite{mahan}, an
electron hopping associated with a time dependent lattice
displacement. As a consequence time retarded electron-phonon
interactions arise in the system yielding a source current which
depends both on time and on the electron path coordinates. This
causes large {\it e-ph} anharmonicities in the equilibrium
thermodynamics of the SSH model \cite{prb04}. Hopping of electrons
from site to site accompanied by a coupling to the lattice
vibration modes is a fundamental process \cite{devreese}
determining the transport \cite{lang} and equilibrium properties
\cite{raedt} of many body systems. A variable range hopping may
introduce some degree of disorder thus affecting the charge
mobility \cite{fish} and the thermodynamic functions.

This paper focusses on this issue, dealing with the
thermodynamical properties of the SSH model in two dimensions and
comparing them with the results obtained in one dimension
\cite{prb03}. Only a few extensions of the SSH Hamiltonian to
higher dimensionality appear in the literature \cite{tang} mainly
concerning the phase diagrams \cite{voo} and the ground state
excitations \cite{figge,ono1,prb02}.  We apply a path integral
method \cite{feynman} which fully accounts for the time retarded
{\it e-ph} interactions and, exploiting the above mentioned
Hamiltonian linear dependence on the atomic displacement, allows
us to derive the electron-phonon source action in two dimensions.
The general formalism is outlined in Section II while the results
are reported on in Section III. The conclusions are drawn in
Section IV.

\section*{II. Path Integral Formalism}

In a square lattice with isotropic nearest neighbors hopping
integral $J$, the SSH Hamiltonian for electrons plus {\it e-ph}
interactions reads:

\begin{eqnarray}
H=\,& & \sum_{r,s}\Bigl[(J_{r,s})_x \bigl(f^{\dag}_{r+1,s} f_{r,s}
+ h.c.\bigr)  \, \nonumber \\ &+& (J_{r,s})_y
\bigl(f^{\dag}_{r,s+1} f_{r,s} + h.c.\bigr) \Bigr] \, \nonumber \\
(J_{r,s})_x=\,& & - {1 \over 2}\bigl[ J - \alpha \Delta u_x \bigr]
\, \nonumber
\\ (J_{r,s})_y=\,& & - {1 \over 2}\bigl[ J - \alpha \Delta u_y \bigr]
\, \nonumber \\ \Delta u_x=\, & & u_x(r+1,s) - {u_{x}(r,s)} \,
\nonumber \\    \Delta u_y=\, & & u_y(r,s+1) - {u_{y}(r,s)} \,
\nonumber \\
\end{eqnarray}

where $\alpha$ is the electron-phonon coupling, ${\bf u}(r,s)$ is
the dimerization coordinate indicating the displacement of the
monomer group on the $(r,s)-$ lattice site, $f^{\dag}_{r,s}$ and
$f_{r,s}$ create and destroy electrons (i.e., $\pi$ band electrons
in polyacetylene). The phonon Hamiltonian is given by a set of 2D
classical harmonic oscillators. The two addenda in (1) deal with
one dimensional {\it e-ph} couplings along the {\it x} and {\it y}
axis respectively, with first neighbors electron hopping. Second
neighbors hopping processes (with overlap integral $J^{(2)}$) may
be accounted for by adding to the Hamiltonian the term $H^{(2)}$
such that

\begin{eqnarray}
& &H^{(2)}=\,(J_{r,s})_{x,y} \bigl(f^{\dag}_{r+1,s+1} f_{r,s} +
h.c.\bigr) \, \nonumber \\ & &(J_{r,s})_{x,y}=\,- {1 \over
2}\Bigl[ J^{(2)} - \alpha \sqrt{ (\Delta u_x)^2 + (\Delta u_y)^2 }
\Bigr] \, \nonumber \\
\end{eqnarray}

The real space Hamiltonian in (1) can be transformed into a time
dependent Hamiltonian \cite{hamann} by introducing the electron
coordinates: i) $\bigl( x(\tau'),y(\tau') \bigr)$ at the $(r,s)$
lattice site, ii) $\bigl( x(\tau),y(\tau') \bigr)$ at the
$(r+1,s)$ lattice site and iii) $\bigl( x(\tau'),y(\tau) \bigr)$
at the $(r,s+1)$ lattice site, respectively. $\tau$ and $\tau'$
vary on the scale of the inverse temperature $\beta$. The spatial
{\it e-ph} correlations contained in (1) are mapped onto the time
axis by changing: $ u_{x,(y)}(r,s) \to u_{x,(y)}(\tau')$,
$u_{x}(r+1,s) \to u_x(\tau)$ and $u_{y}(r,s+1) \to u_y(\tau)$. Now
we set $\tau'=\,0$, $\bigl( x(\tau'),y(\tau') \bigr) \equiv
(0,0)$, $\bigl( u_x(\tau'),u_y(\tau') \bigr) \equiv (0,0)$.
Accordingly, (1) transforms into the time dependent Hamiltonian:

\begin{eqnarray}
H(\tau)&=&\,J_x(\tau) \Bigl(f^{\dag}(x(\tau),0)f(0,0) + h.c.
\Bigr)\, \nonumber
\\ &+& J_y(\tau) \Bigl(f^{\dag}(0,y(\tau))f(0,0) + h.c. \Bigr)
\, \nonumber
\\
J_{x}(\tau)&=&\,- {1 \over 2}\bigl[J - \alpha u_{x}(\tau) \bigr]
\, \nonumber
\\
J_{y}(\tau)&=&\,- {1 \over 2}\bigl[J - \alpha u_{y}(\tau) \bigr]
\, \nonumber
\\
\end{eqnarray}

While the ground state of the 1D SSH Hamiltonian is twofold
degenerate, the degree of phase degeneracy is believed to be much
higher in 2D \cite{ono2} as many lattice distortion modes
contribute to open the gap at the Fermi surface. Nonetheless, as
in 1D, these phases are connected by localized and nonlinear
excitations, the soliton solutions. Thus, also in 2D both electron
hopping between solitons \cite{kivel} and thermal excitation of
electrons to band states may take place within the model. These
features are accounted for by the time dependent version of the
Hamiltonian.

As $\tau$ varies continuously on the $\beta$ scale and the
$\tau$-dependent displacement fields are continuous variables
(whose amplitudes are in principle unbound in the path integral),
long range hopping processes are automatically included in
$H(\tau)$ which is therefore more general than the real space SSH
Hamiltonian in (1) (and (2)). Thus by means of the path integral
formalism we look at the low temperature thermodynamical behavior
both in 1D and 2D searching for those features which may be
ascribable to some local disorder related to the variable range of
the hopping processes.

The semiclassical nature of the model is evident from (3) in which
quantum mechanical degrees of freedom interact with the classical
variables $u_{x (y)}(\tau)$. Averaging the electron operators over
the ground state we obtain the time dependent semiclassical energy
per lattice site $N$:

\begin{eqnarray}
& &{{<H(\tau)>} \over N}=\,J_{x}(\tau)P\bigl(J, \tau,
x(\tau)\bigr)  + J_{y}(\tau) P\bigl(J, \tau, y(\tau)\bigr) \,
\nonumber \\ & &P\bigl(J, \tau, {\bf v}(\tau)\bigr)=\, {1 \over
{\pi^2}}\int d{\bf k} \cos[{\bf k \cdot v}(\tau)]
\cosh(\epsilon_{\bf k} \tau) n_F(\epsilon_{\bf k})  \, \nonumber
\\
\end{eqnarray}

with ${\bf v}(\tau)=\,(x(\tau),0)$ and ${\bf
v}(\tau)=\,(0,y(\tau))$ in the first and second addendum
respectively. $\epsilon_{\bf k}=\, -J \sum_{i=x,y} \cos(k_i)$ is
the electron dispersion relation and $n_F$ is the Fermi function.
Eq.(4) can be rewritten in a way suitable to the path integral
approach by defining

\begin{eqnarray}
& &{{<H(\tau)>} \over N}=\, V\bigl({x}(\tau)\bigr) +
V\bigl({y}(\tau)\bigr) + {\bf u}(\tau) \cdot {\bf j}({\bf
v}(\tau)) \, \nonumber
\\ & &V\bigl({x}(\tau)\bigr)=\,-J P\bigl(J, \tau, {x}(\tau)\bigr)
\, \nonumber
\\ & &V\bigl({y}(\tau)\bigr)=\,-J P\bigl(J, \tau, {y}(\tau)\bigr)
\, \nonumber
\\& &{\bf j}({\bf v}(\tau))= -\alpha P\bigl(J, \tau,
{\bf v}(\tau)\bigr) \, \nonumber \\ & &{\bf
u}(\tau)=\,\bigl(u_{x}(\tau),u_{y}(\tau)\bigr) \, \nonumber \\
\end{eqnarray}

$V\bigl({x}(\tau)\bigr)$ and $V\bigl(y(\tau)\bigr)$ are the
effective terms accounting for the $\tau$ dependent electronic
hopping while ${\bf j}({\bf v}(\tau))$ is interpreted as the
external source \cite{kleinert} current for the oscillator field
${\bf u}(\tau)$. Averaging the electrons over the ground state we
neglect the fermion-fermion correlations \cite{hirsch}. This
approximation however is not expected to affect substantially the
following calculations. Taking a bath of $\bar N$ 2D oscillators,
we generally write the SSH electron path integral, ${\bf
\zeta}(\tau) \equiv \,\bigl({x}(\tau),{y}(\tau)\bigr)$,  as:

\begin{eqnarray}
<{\bf \zeta}(\beta)&|&{\bf \zeta}(0)>=\,\prod_{i=1}^{\bar N} \int
D{\bf u}_i(\tau)  \int D{\bf \zeta}(\tau)  \, \nonumber
\\
&\cdot& exp\Biggl[- \int_0^{\beta}d\tau \sum_{i=1}^{\bar N} {M
\over 2} \biggl( \dot{\bf u}_i^2(\tau) + \omega_i^2 {\bf
u}_i^2(\tau) \biggr)\Biggr] \, \nonumber
\\
&\cdot& exp\Biggl[- \int_0^{\beta}d\tau \biggl({m \over 2}
\dot{\bf \zeta}^2(\tau) + V\bigl({x}(\tau)\bigr) +
V\bigl({y}(\tau)\bigr) \biggr)\, \nonumber
\\ &+& \sum_{i=1}^{\bar N} {\bf
u}_i(\tau) \cdot {\bf j}({\bf v}(\tau)) \Biggr] \, \nonumber
\\
\end{eqnarray}

where $m$ is the electron mass, $M$ is the atomic mass and
$\omega_i$ is the oscillator frequency.  As a main feature we
notice that the interacting energy is linear in the atomic
displacement field. Then, the electronic path integral can be
derived after integrating out the oscillator degrees of freedom
which are decoupled along the $x$ and $y$ axis. Thus, we get:

\begin{eqnarray}
& &<{\bf \zeta}(\beta)|{\bf \zeta}(0)>=\,\prod_{i=1}^{\bar N} Z_i
\Biggl[ \int D{x}(\tau) \, \nonumber
\\ &\cdot& exp\Bigl[- \int_0^{\beta}d\tau \biggl({m \over 2}
\dot{x}^2(\tau) + V\bigl({x}(\tau)\bigr) \biggr) - {1 \over \hbar}
A({x}(\tau)) \Bigr] \Biggr]^2 , \, \nonumber \\ & &A({
x}(\tau))=\, -{{\hbar^2} \over {4M}}\sum_{i=1}^{\bar N} {1 \over
{\hbar \omega_i \sinh(\hbar\omega_i\beta/2)}} \, \nonumber \\
&\cdot& \int_0^{\beta} d\tau j({x}(\tau)) \int_0^{\beta}d{\tau''}
\cosh\Bigl(\hbar\omega_i \bigl( |\tau - {\tau''}| - \beta/2 \bigr)
\Bigr) j({x}({\tau''})), \, \nonumber
\\ & &Z_i=\, \Bigl[ {1 \over {2\sinh(\hbar\omega_i\beta/2)}}
\Bigr]^2 \, \nonumber
\\
\end{eqnarray}

and the 2D electron path integral is obtained after squaring the
sum over one dimensional electron paths. This permits to reduce
the computational problem which is nonetheless highly time
consuming particularly in the low temperature limit. Note infact
that the source current $j({x}(\tau))$ requires integration over
the 2D Brillouin Zone (BZ) according to (4) and (5) and this
occurs for any choice of the electron path coordinates. The
quadratic (in the coupling $\alpha$) source action $A({x}(\tau))$
is time retarded as the particle moving through the lattice drags
the excitations of the oscillator fields which take a time to
readjust to the electron motion. When the interaction is
sufficiently strong the conditions for polaron formation
\cite{sto} may be fulfilled in the system according to the degree
of adiabaticity \cite{brown}. However the present path integral
description is valid independently of the existence of polarons as
it applies also to the weak coupling regime. Assuming periodic
conditions ${x}(\tau)=\, {x}(\tau + \beta)$, the particle paths
can be expanded in Fourier components

\begin{eqnarray}
& &{x}(\tau)=\,{x}_o + \sum_{n=1}^\infty 2\Bigl(\Re {x}_n \cos(
\omega_n \tau) - \Im {x}_n \sin( \omega_n \tau) \Bigr)\, \nonumber
\\ & &\omega_n=\,2\pi n/\beta
\end{eqnarray}

and the open ends integral over the paths $\int D{x}(\tau)$
transforms into the measure of integration $\oint D{x}(\tau)$.
Taking:

\begin{equation}
\oint D{x}(\tau)\equiv \int_{-\infty}^{\infty}{{d{x}_o} \over
{\bigl( 2\pi\hbar^2/mK_BT \bigr)^{(1/2)}}} \prod_{n=1}^{\infty}
\Biggl[{{\int_{-\infty}^{\infty} d\Re {x}_n
\int_{-\infty}^{\infty} d\Im {x}_n} \over {\bigl( \pi \hbar^2
K_BT/m\omega_n^2 \bigr)}} \Biggr]
\end{equation}

which normalizes the kinetic energy term in (7), we proceed to
integrate (7) in order to derive the full partition function of
the system versus temperature.

\section*{III. Thermodynamics in 1D and 2D}

Taking a dimensionless path (in units of the lattice constant
$a=\,1 \AA$) with: ${a}_n \equiv 2 \Re {x}_n $ and ${b}_n \equiv
-2 \Im {x}_n $ the functional measure in (9) can be rewritten as:

\begin{eqnarray}
\oint D{x}(\tau) &\approx& {{2^{1/2}} \over {{\bigl(2 \lambda_m
\bigr)}^{(2N_p+1)}}} \prod_{n=1}^{N_p} (2 \pi n)^{2}
\int_{-\Lambda}^{\Lambda}d {x}_o  \, \nonumber \\ & \cdot
&\int_{-2\Lambda}^{2\Lambda}d {a}_n \int_{-2\Lambda}^{2\Lambda}d
{b}_n \, \nonumber \\ \lambda_m &=&\, \sqrt{{ \pi \hbar^2} \over
{m K_BT}}
\end{eqnarray}

where $N_p$ is the cutoff on the Fourier components in (8). At a
given temperature and for a given set of particle path
coefficients we first carry out the momentum integrations required
by (4) summing over 1600 points in the reduced 2D BZ. This
prepares the source current $j({x}(\tau))$ to be used in the
double time integration which yields, according to (7), the source
action $A({x}(\tau))$. Afterwards we sum the exponential of the
total action over an appropriate set of particle paths.
Numerically stable results are achieved by tuning the number of
pairs $N_p$ in (10), the cutoff ($\Lambda$) on the integration
range of the Fourier coefficients in (10) and the related number
of points ($N_\Lambda$) in the measure of integration. $N_p=\,2$
suffices in the Fourier expansion of the path. The argument to set
the cutoff $\Lambda$ on the integration range is based on the fact
that the functional measure normalizes the kinetic term in (7):

\begin{equation} \oint D{\bf x}(\tau) exp\Biggl[-
\int_0^{\beta}d\tau {m \over 2} \dot{\bf x}^2(\tau) \Biggr] =\,1
\end{equation}

From (11) we get, $\Lambda \sim 3 \lambda_m / \sqrt{2 \pi^3}$
hence, $\Lambda$ scales versus temperature as $\Lambda \propto
1/\sqrt{T}$. Computing (7) with (10) we find that infact a smaller
cutoff, $\Lambda \sim \lambda_m /(10 \sqrt{2 \pi^3})$, guarantees
convergence in the interacting partition function. Accordingly the
number of points in each integration range has to increase by
decreasing $T$ as $N_\Lambda \propto 1/\sqrt{T}$. In the 2D
problem, $N_\Lambda \sim 35 /\sqrt{T}$ points for each Fourier
coefficient are required in (10). Then we carry out the total 2D
path integral by evaluating the contribution of $(N_\Lambda +
1)^{2N_p + 1}$ paths selecting the integer part of $N_\Lambda$ at
any temperature. Computation of the second order derivatives of
the free energy in the range $T\in [1,301]K$, with a spacing of
3K, takes 55hours and 15 minutes on a Pentium 4.

The physical input parameters characterizing both the 1D and 2D
models are: the bare hopping integral $J$, the oscillator
frequencies $\omega_i$ and the effective coupling $\chi=\,
\alpha^2 \hbar^2 /M$ (in units $meV^3$). We take here a narrow
band system with $J=\,100meV$. The crossover between weak and
strong {\it e-ph} coupling is given in our model by $\chi_c \sim
\, \pi J \hbar^2 \omega^2_{max}/64$ \cite{lu}, where
$\omega_{max}$ is the largest energy in the oscillators bath.
Let's assume a bath of $\bar N=\,10$ low energy oscillators with
$\hbar \omega_{i} \sim \,20meV$ thus setting the crossover value
at $\chi_c \sim \,2000meV^3$.  A higher number of oscillators in
the same energy range would not significantly modify the results
hereafter presented whereas lower $\omega_i$ values would yield a
larger contribution to the phonon partition function mainly at low
$T$. In Fig.1, a comparison between the 1D and the 2D free
energies is presented for two values of $\chi$, one lying in the
weak and one in the strong {\it e-ph} coupling regime. The
oscillator free energies $F_{ph}$ are plotted separately while the
free energies arising from the total action in (7), shortly termed
$F_{source}$, are due to the purely electronic plus {\it e-ph}
contributions. In general, the 2D free energies have a larger
gradient (versus temperature) than the corresponding 1D terms. The
$F_{ph}$ lie above the $F_{source}$ both in 1D and 2D because of
the choice of the $\hbar \omega_{i}$; lowering the energy of some
oscillator modes would produce a crossing point between $F_{ph}$
and $F_{source}$ with temperature location depending on the value
of $\chi$. Note that the $F_{source}$ plots have a positive
temperature derivative in the low temperature regime and this
feature is more pronounced in 2D. Infact, at low $T$, the source
action is dominated by the hopping potential
$V\bigl(x(\tau)\bigr)$ while, at increasing $T$, the {\it e-ph}
effects become progressively more important (as the bifurcation
between the $\chi_1$ and $\chi_2$ curves shows) and the
$F_{source}$ get a negative derivative. In 2D, the weight of the
$V\bigl(x(\tau)\bigr)$ term is larger because there is a higher
hopping probability. This physical expectation is taken into
account by the path integral method. At any temperature, we
monitor the ensamble of relevant particle paths over which the
hopping potential is evaluated. For a selected set of Fourier
components in (10) the hopping decreases by lowering $T$ but its
value is still significant at low $T$. Considered that: i) the
total action is obtained after a $d\tau$ integration of
$V\bigl(x(\tau)\bigr)$ and ii) the $d\tau$ integration range is
larger at lower temperatures, we explain why the overall hopping
potential contribution to the total action is responsible for the
anomalous free energy behavior at low T.

In Fig.2, the heat capacity contributions  due to the oscillators
($C_{ph}$) and electrons plus {\it e-ph} coupling ($C_{sou}$) are
reported on. The values are normalized over $\bar N$. The
dimensionality effects are seen to be large and, for a given
dimensionality, the role of the {\it e-ph} interactions is
magnified at increasing $T$. The total heat capacity ($C_{ph}$ +
$C_{sou}$) over $T$ ratios are plotted in Fig.3 in the low $T$
regime to emphasize the presence of an anomalous upturn which
appears at $T \preceq 10K$ in 1D and $T \preceq 20K$ in 2D. This
feature in the heat capacity linear coefficient is ultimately
related to the above discussed overall hopping potential effects
while the strength of the coupling $\chi$ has a minor role in the
low $T$ limit.

Finally we emphasize that the degeneracy of the ground state
scarcely affects the finite temperature equilibrium properties.
This can be shown in 1D by treating the SSH Hamiltonian in mean
field approximation with a staggered dimerization amplitude which
yields the size of the lattice distortion. By diagonalizing the
Hamiltonian in momentum space we see that the competition between
electron and lattice energy determines the contribution to the
partition function and shifts downwards the free energy versus
temperature. Hence, the heat capacity due to the degeneracy is
very small (in comparison with the values given in Fig.2) and
strictly linear in $T$. Accordingly, the peculiar and sizeable
upturn in the heat capacity over T ratio does not depend on the
nature of the ground state. The inclusion of the quantum lattice
fluctuations beyond the mean field approximation is expected to
reinforce our conclusion as the fluctuations tend to suppress the
long range order associated with the dimerized structure
\cite{pere}.

\section*{IV. Conclusions}

We have investigated the thermodynamics of a semiclassical
Su-Schrieffer-Heeger (SSH) Hamiltonian extended to two dimensions
in which the lattice displacements are taken as classical
variables. By mapping the real space interactions onto the time
($\tau$) scale we deal with an Hamiltonian accounting for a
variable range of the electronic transfer integral. Such a model
can be suitably treated in the path integral formalism as both the
electron and lattice degrees of freedom are continuous functions
of $\tau$ which varies up to the inverse temperature. Although we
are thus assuming a continuum model the present path integral
calculation is valid for any range of {\it e-ph} coupling and for
any value of physical parameters.

The two dimensional path integral for the SSH particle in a bath
of harmonic oscillators has been derived by summing over a set of
electron paths that becomes increasingly large at decreasing
temperatures. Thus, the computation of the total action and of the
free energy derivatives is highly time consuming in the low $T$
regime where the quantum mechanical nature of the electronic
subsystem is fully accounted for. At low $T$, the electron hopping
integral largely shapes the free energy which displays an
anomalous positive derivative more pronounced in 2D than in 1D.
This behavior is mirrored in the upturn of the {\it heat capacity
over T} ratio. For a given set of input parameters characterizing
electron band, oscillator energies and {\it e-ph} couplings, we
find that the upturn shows up at $T \preceq 20K$ in the square
lattice and at $T \preceq 10K$ in the linear chain. Also the size
of the upturn is larger in two dimensions consistently with the
fact that the 2D hopping potential contribution to the total
action is higher than in 1D. The presented results have been
obtained assuming a narrow electron band ($J \sim 100meV$) but the
upturn in the {\it heat capacity over T} ratio appears also in the
case of wide band systems.

The degenerate ground state of the SSH model is not responsible
for the anomalous behavior of the heat capacity over $T$ ratio.
Infact, closing the linear chain in a ring geometry, one can
easily diagonalize the 1D Hamiltonian in mean field approximation
and show that the contribution to the total heat capacity, due to
the two fold degeneracy of the ground state, is small and linear
versus temperature. A quantum description of the lattice
displacement fields should further reduce the role of the
dimerized ground state as the fluctuations tend to suppress the
order parameter. Similar conclusions are expected to hold in the
2D case where, however, the degeneracy of the ground state is
manyfold.

In conclusion, while in the low $T$ regime the thermodynamics is
essentially governed by the electronic term of the Hamiltonian, at
increasing $T$ the strength of {\it e-ph} coupling progressively
determines the slope of the free energy and heat capacity. Our
results suggest that a glassy like thermodynamical behavior is a
peculiar feature of Hamiltonian models with a variable range for
the electronic hopping. Such a behavior is more likely to occur in
higher dimensionality  as the structure is more closely packed
and, accordingly, the hopping probability is larger.

\begin{figure}
\vspace*{7truecm} \caption{Colour online. Phonon  ($F_{ph}$) and
Source Term ($F_{sou}$) contributions to the 1D and 2D free
energies for two values of the effective coupling $\chi$:
$\chi_1=\,1440 meV^3$ (weak {\it e-ph} coupling) $\chi_2=\,2560
meV^3$ (strong {\it e-ph} coupling).}
\end{figure}

\begin{figure}
\vspace*{7truecm} \caption{Colour online. Phonon and Source Term
contributions (normalized over the number of oscillators) to the
1D and 2D heat capacities for the same parameters as in Fig.1. The
oscillator heat capacities are also plotted.}
\end{figure}

\begin{figure}
\vspace*{7truecm} \caption{Colour online. Total heat capacity over
temperature for the same parameters as in Fig.1. }
\end{figure}


\begin{references}
\bibitem{peierls}
R.E.Peierls, Quantum Theory of Solids, Clarendon (Oxford) 1955.
\bibitem{lu}
Yu Lu, {\it Solitons and Polarons in Conducting Polymers} World
Scientific, Singapore (1988).
\bibitem{ssh}
W.P.Su, J.R.Schrieffer, A.J.Heeger,  Phys.\ Rev.\ Lett. {\bf 42},
1698 (1979).
\bibitem{ssh1}
A.J. Heeger, S.Kivelson, J.R.Schrieffer,
W.-P.Su, Rev.\ Mod.\ Phys.\ {\bf 60}, 781 (1988).
\bibitem{mahan}
G.D.Mahan, {\it Many Particle Physics}, Plenum Press, N.Y. (1981).
\bibitem{prb04}
M.Zoli, Phys.Rev.B {\bf 70}, 184301 (2004).
\bibitem{devreese}
J.T.Devreese, {\it Polarons} in Encyclopedia of Applied Physics
(VCH Publishers, NY) {\bf 14}, 383 (1996).
\bibitem{lang}
I.J.Lang, Yu.A.Firsov, Sov. Phys. JETP {\bf 16}, 1301 (1963).
\bibitem{raedt}
H.De Raedt, A.Lagendijk, Phys.\ Rev.\ B {\bf 27}, 6097 (1983).
\bibitem{fish}
I.I.Fishchuk, A.Kadashchuk, H.B\"{a}ssler, S.Ne\v{s}purek, Phys.\
Rev.\ B {\bf 67}, 224303 (2003).
\bibitem{prb03}
M.Zoli, Phys.Rev.B {\bf 67}, 195102 (2003).
\bibitem{tang}
S.Tang, J.E.Hirsch,  Phys.\ Rev.\ B {\bf 37}, 9546 (1988).
\bibitem{voo}
K.K.Voo, C.Y.Mou, Cond-Mat/0309298
\bibitem{figge}
M.T.Figge, M.Mostovoy, J.Knoester, Phys.\ Rev.\ Lett. {\bf 86},
4572 (2001).
\bibitem{ono1}
N.Miyasaka,Y.Ono, J.Phys.Soc.Jpn. {\bf 70}, 2968 (2001).
\bibitem{prb02}
M.Zoli, Phys.\ Rev.\ B {\bf 66}, 012303 (2002).
\bibitem{feynman}
R.P.Feynman, Phys. Rev. {\bf 97}, 660 (1955).
\bibitem{hamann}
D.R.Hamann, Phys.\ Rev.\ B {\bf 2}, 1373 (1970).
\bibitem{ono2}
Y.Ono, T.Hamano, J.Phys.Soc.Jpn. {\bf 69}, 1769 (2000).
\bibitem{kivel}
S.Kivelson, Phys.\ Rev.\ Lett. {\bf 46}, 1344 (1981).
\bibitem{kleinert}
H.Kleinert, {\it Path Integrals in Quantum Mechanics, Statistics
and Polymer Physycs} World Scientific Publishing, Singapore
(1995).
\bibitem{hirsch}
J.E.Hirsch, Phys.\ Rev.\ Lett. {\bf 51}, 296 (1983).
\bibitem{sto}
V.M.Stojanovi\'{c}, P.A.Bobbert, M.A.J.Michels, Phys.\ Rev.\ B
{\bf 69}, 144302 (2004).
\bibitem{brown}
D.W.Brown, K.Lindenberg, Y.Zhao, J.Chem.Phys. {\bf 107}, 3179
(1997).
\bibitem{pere}
V.Perebeinos, P.B.Allen, J.Napolitano, Solid State Commun. {\bf
118}, 215 (2001).

\end{references}
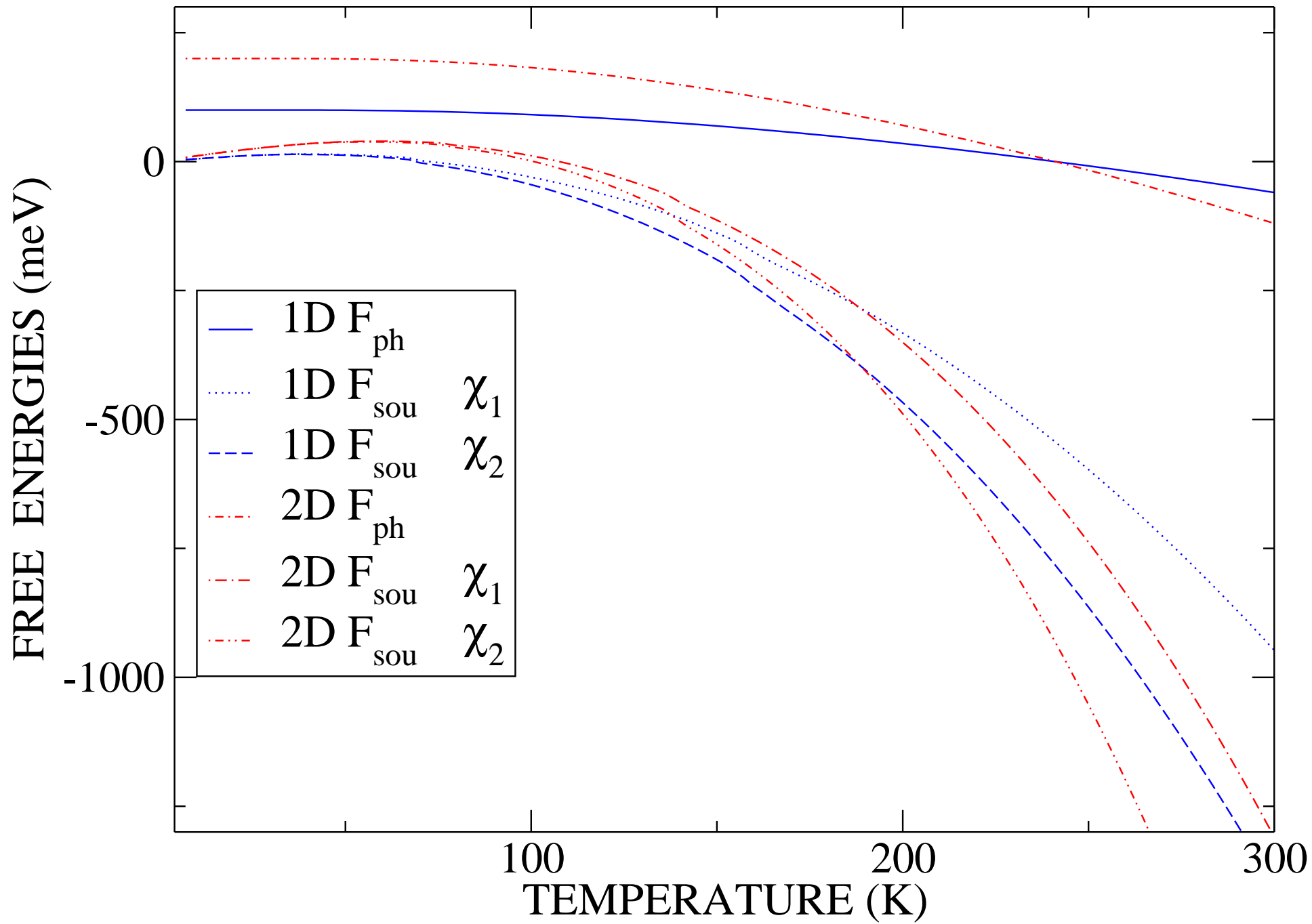
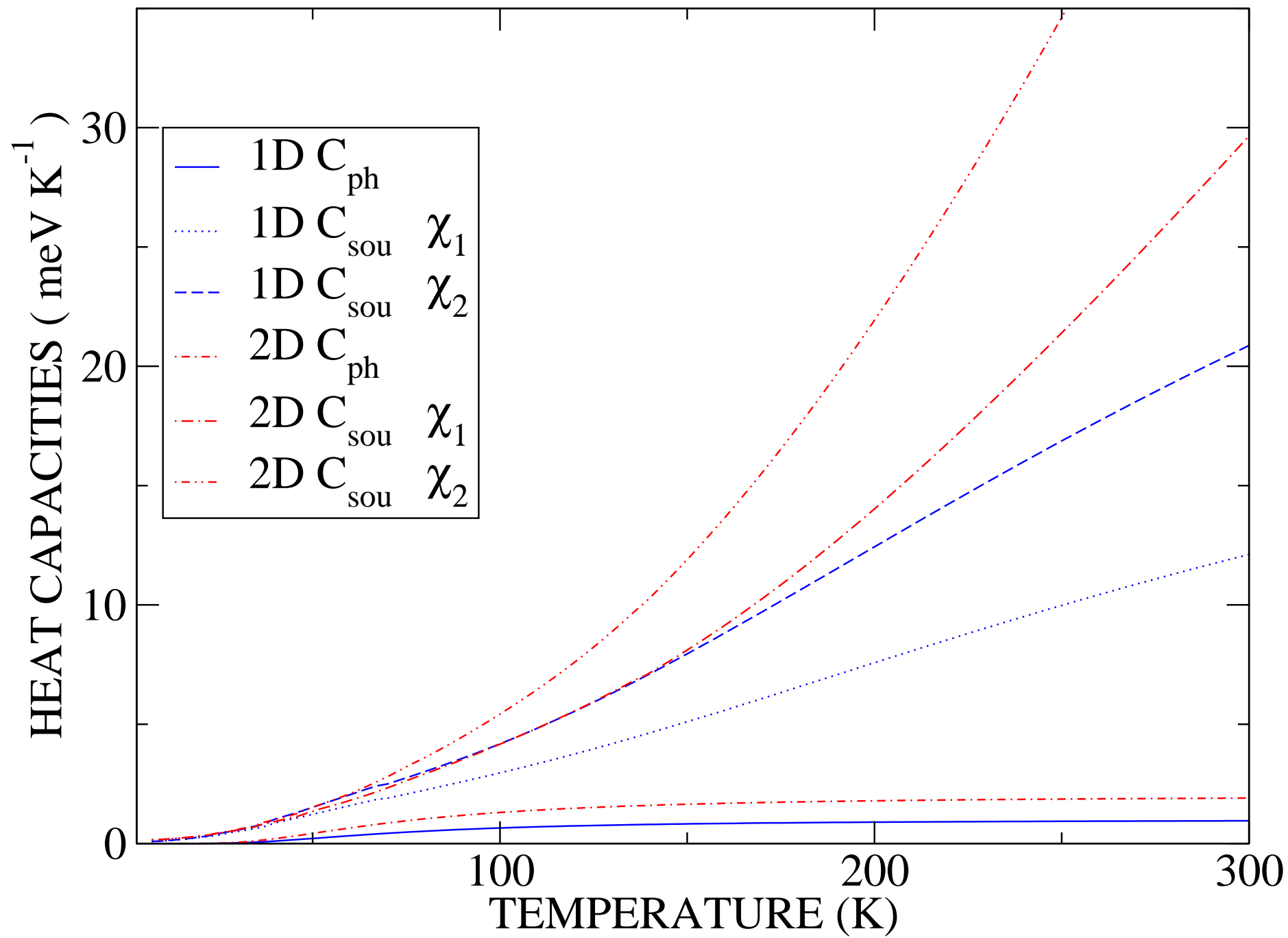
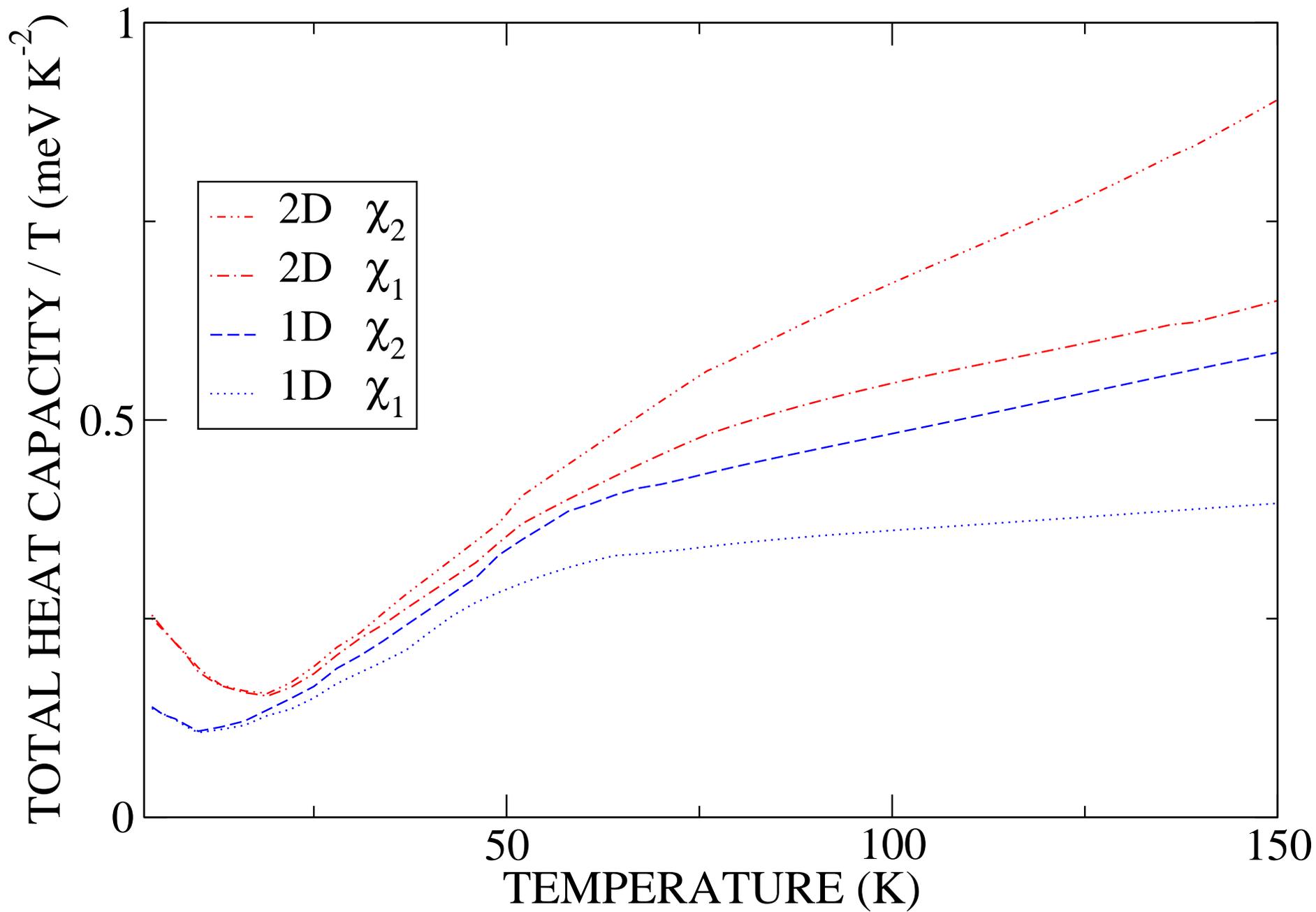
\end{document}